\documentstyle[pre,aps,preprint]{revtex}

\def\bh{{\rm\scriptscriptstyle BH}}
\def\gas{{\rm gas}}
\def\nuc{{\rm nuc}}
\def\rhoin{\rho_{\rm in}}
\def\rhout{\rho_{\rm out}}

\begin{document}

\draft
\tighten

\title{On the Observability of Quantum Information Radiated
       from a Black Hole}

\author{
Mark Srednicki\footnote{E--mail: \tt mark@physics.ucsb.edu}
       }
\address{Department of Physics, University of California,
         Santa Barbara, CA 93106 }

\maketitle

\begin{abstract}
We propose a resolution to the black-hole information-loss
paradox: in one formulation of physical theory, information
is preserved and macroscopic causality is violated; in another,
causality is preserved and pure states evolve to mixed states.
However, no experiments can be performed 
that would distinguish these two descriptions.  We explain how
this could work in practice; a key ingredient is the suggested
quantum-chaotic nature of black holes.
\end{abstract}

\pacs{04.70.Dy, 11.25.-w}

\section{Introduction}
\label{1}

Twenty-seven years ago, Hawking proposed that black-hole evaporation
required a modification of quantum mechanics: pure states had to
be able to evolve into mixed states \cite{h1}.  
Otherwise, there would have to be nonlocal effects operating
on macroscopically large scales, in regions of low space-time 
curvature \cite{nice}; this was
(and remains) an anathema to devotees of Einstein's theory.
On the other hand, the evolution of
pure to mixed states was equally abhorrent to devotees of
the theory of Schr\"odinger, Heisenberg, and Dirac.  
Consequently one's position
on this key issue of fundamental physics has tended to be
sociologically determined, though there are some notable
exceptions.

The discovery that string theory allowed one to count black-hole
microstates \cite{sv}, 
and get a number that agreed with the one predicted
by the Bekenstein-Hawking entropy \cite{jdb,h2,h3}, gave a strong boost 
to those on the side of quantum mechanics: string theory is 
explicitly a quantum mechanical theory, 
and it clearly violates field theoretic
notions of locality.  More recently, the discrete
light-cone quantization scheme (DLCQ, also known as the ``matrix
model'' \cite{matrix}), and the equivalence of
string theory in anti-de-Sitter space to a conformal field theory
in a lower dimension (AdS/CFT \cite{adscft}), give us, in principle,
explicit hamiltonian descriptions of black-hole formation
and decay.  This would seem to settle the issue decisively in
favor of quantum mechanics, at least for string-theoretic
descriptions of quantum gravity.

However, in all the years of debate, it has been implicitly assumed
that the question of whether pure states evolve to mixed states
could, in principle, be answered experimentally by performing 
appropriate measurements.

In this paper, I wish to call this into question.  I will suggest
a physical picture of the quantum states of black holes such that,
even if the string-theoretic scenario outlined above is correct,
experiments would still show apparent information loss during black
hole evaporation.  This then further suggests that there may be a dual
or complementary description of this process in which pure states
evolve to mixed states, but macroscopic causality is preserved.
These two descriptions would be experimentally indistinguishable.

\section{Black-Hole Resonances in the S-Matrix}
\label{2}

Black holes are unstable, and decay by emitting Hawking 
radiation \cite{h2}.
If quantum mechanics is unmodified, black holes will show up
as resonances in the $S$-matrix \cite{smatrix} for scattering of 
stable particles \cite{foot}.
In the complex $s$ plane (where the Mandelstam variable $s$ is
the total center-of-mass energy squared), there will be poles at
\begin{equation}
s_n = M_n^2 - i \Gamma_n ,
\label{sn}
\end{equation}
with
\begin{equation}
M_{n+1} - M_n = C_n \exp(-4\pi M_n^2).
\label{mn}
\end{equation}
Here $\Gamma_n$ and $C_n$ are numbers of order one in Planck
units ($\hbar=c=G=k=1$).
These formulas are appropriate for Schwarzschild black holes in
3+1 dimensional spacetime.  
Eq.~(\ref{mn}) follows from the density of
black-hole states $e^S$, where $S=4\pi M^2$ is the
Bekenstein-Hawking entropy.  
Eq.~(\ref{sn}) follows from the Hawking decay rate 
\begin{equation}
{dM\over dt} = - 4\pi R^2 \sigma T^4,
\label{dmdt}
\end{equation}
where 
$T=1/8\pi M$ is the Hawking temperature,
$R=2M$ is the Schwarzschild radius, and 
$\sigma = g_s\pi^2/30$ is the Stefan-Boltzmann constant
($g_s$ is the number of effectively massless particle species
at temperature $T$). 
Dividing $dM/dt$ by the typical energy $T$ of an
emitted quantum gives us the width of the resonance in the
energy plane, and multiplying by $M$ gives us the width
$\Gamma_n$ in the $s$ (energy-squared) plane 
(assuming $\Gamma_n \ll M_n$).  We conjecture that the
numbers $\Gamma_n$ and $C_n$ vary erratically with $n$;
we will elaborate on this in Section~\ref{3}.

Eq.~(\ref{mn}) is a highly unusual behavior for a set
of resonances.  Not only does their density increase
rapidly with $n$, but the series never terminates.
Other known systems (such as complex atoms or nuclei) 
have dense resonances, but eventually there is a threshold 
(for, e.g., ionization or fission), and above this threshold
the poles are replaced by a cut.  This is not supposed to happen
for black holes, which can be arbitrarily large.
This makes the very definition of the $S$-matrix
problematic.  To define the $S$-matrix, one typically
considers widely separated (and therefore noninteracting)
wave packets for individual particles that come together
in an interaction region that is bounded in space and 
time.  It is assumed that this can be done for
arbitrarily large energy, but here large energy
implies a large black hole, and thus the size of
the interaction region grows with energy.

To regularize this problem, let us compactify
the three dimensions of space onto a torus.
We will consider this system at all energies,
and try to deduce the properties of the
$S$-matrix that could be extracted from it.

\section{Black Holes in a Box}
\label{3}

Consider putting energy $E$ into our toroidal box
of linear size $L$.  We will consider two possible
generic forms for this energy: a gas of massless
particles, and a Schwarzschild black hole.  
A gas of massless particles with energy
$E_\gas$ in a volume $L^3$ has entropy
$S_\gas \sim (LE_\gas)^{3/4}$.  A Schwarzschild black
hole of mass $M$ has entropy $S_\bh = 4\pi M^2$.
If we maximize the total entropy $S_\gas + S_\bh$
subject to the constraint $E_\gas + M = E$ (this
is equivalent to requiring the gas and the black
hole to have the same temperature), we find 
$M^4 E_\gas \sim L^3$.  Dividing both sides by $E^5$,
we see that the left hand-side is always less than
one, which means there is a solution only if 
$E > L^{3/5}$.  In this case we find
$M=(1-\varepsilon)E$ and $E_\gas = \varepsilon E$,
where $\varepsilon \sim L^3/E^5$.
If $E < L^{3/5}$, then the entropy is maximized
by putting all of the energy into the gas.  In this
regime, the probability of finding a black hole of mass
$M$ is governed by the Boltzmann factor $\exp(-M/T_\gas)$,
where $T_\gas=(\partial S_\gas/\partial E)^{-1}$ is the
gas temperature.

We now assume that some sort of hamiltonian description of this
black-hole-in-a-box is available, analogous to the hamiltionians
provided by DLCQ or AdS/CFT.
The thermodynamic analysis in the previous paragraph
made implicit use of the microcanonical ensemble: the defining
thermodynamic relation is entropy as a function of energy and
volume.  For classical systems, microcanonical averages involve
integrating over a constant-energy surface in phase space.
For quantum systems, one traditionally sums over 
energy eigenstates in some ``small'' energy range.  
A natural question is, how small can this energy range
be? Or, equivalently, how many energy eigenstates must we
use for the microcanonical averaging?
For interacting many-body quantum systems that behave
chaotically, the answer is now believed to be {\it one}\cite{qc}.  
That is, individual energy eigenstates have the properties 
of a state of thermal equilibrium.  
(To motivate this for the reader who does not want to go 
through the quantum-chaos literature, note that energy
eigenstates are time independent; thus a system in one 
of them does not evolve.  If an isolated complex system 
of many interacting particles does not evolve, then, 
according to standard thermodynamic reasoning, it must 
be in a state of thermal equilibrium.  This argument was 
first made for black holes in \cite{rama}.)  
In the context of black-hole thermodynamics, {\it thermal\/}
means {\it classical\/}.  Thus, for example, if we can
identify an operator $\hat R^\mu{}_{\nu\rho\sigma}(x)$
corresponding to the curvature tensor, and $|\alpha\rangle$
is an energy eigenstate in the black-hole regime $E \gg L^{3/5}$,
then $\langle\alpha|\hat R^\mu{}_{\nu\rho\sigma}(x)|\alpha\rangle$
should be equal to the Schwarzschild curvature, up to corrections
of order $\varepsilon \sim L^3/E^5$. 

This means that if we put energy $E\gg L^{3/5}$ into a box of
linear size $L$, we should find energy eigenstates that look
very much like black holes of mass $E$.  Note that we must have
$R=2E \ll L$ for the black hole to fit into the box, but this
is not a problem for large $L$.  If we take
$E=L^{4/5}$, for example, then we obviously have 
$L^{3/5} \ll E \ll L$ when $L$ is large.  
In this case,
for an energy equivalent to one solar mass,
the Schwarzschild radius is 3$\,$km,
and $L$ is nearly a billion times larger.

The emergence of black holes out of a gas in a box
at high energy is
due to the negative specific heat of the black hole.
Consider what would happen if we put, say, a uranium nucleus 
in the box instead.  A nucleus with energy $E$ (we work 
in units where $E=0$ is the energy of the ground state,
$E=1$ is the energy of the first excited state, and
$\hbar=c=1$) has entropy (logarithm of the density of 
quantum states) $S_\nuc \sim \sqrt{E}$.
Suppose we give our nucleus energy $E$; it will decay
by emission of various particles, which we will suppose
ultimately form a gas of entropy 
$S_\gas \sim (LE_\gas)^{3/4}$; the nucleus
will then have energy $E_\nuc = E-E_\gas$.
To determine $E_\nuc$ and $E_\gas$,
we again minimize the total entropy 
at constant total energy.
This time we find that the system is dominated by the 
nucleus only at {\it low\/} energy, $E < L^{-3}$.
In our units, the size of the nucleus
is roughly one, so we cannot have a small ($L < 1$) box.
For a large ($L\gg 1$) box, we are limited to such low energies
that we can see only the ground state of the nucleus ($E \ll 1$).
Thus, unlike a black hole, we cannot study the 
quantum states of a nucleus by putting it in a box and letting
it come to thermal equilibrium with its decay products.

We now make note another result from quantum chaos theory.
While individual energy eigenstates give thermal 
expectation values to simple observables, the wave functions
of these states are very complex and easily disturbed
by small perturbations.  Correspondingly, the precise
energy eigenvalues are also highly sensitive to perturbations.
Thus, if we were to change the shape
of our box (or, equivalently, the modular parameters of the 
compactification torus), we would expect the thermal and 
classical aspects of these states to be unchanged (so that 
$\langle\alpha|\hat R^\mu{}_{\nu\rho\sigma}(x)|\alpha\rangle$
would still have the Schwarzschild value),
but the detailed energy eigenvalues and eigenfunctionals
would be different.  This is true even though the walls of
the box can be arbitrarily many Schwarzschild radii away.

The combination of these two results (dominance of the black-hole
states at high energy, and sensitivity of the energy eigenvalues
to large-scale boundary conditions) leads us to conjecture that
the locations of the black-hole resonance
poles in the $S$-Matrix, eq.~(\ref{sn}), will also be sensitive
to the large-scale boundary conditions.  

Another way to see how this sensitivity could arise is to
consider small changes in local fields (such as the metric).
In the language of Feynman diagrams, these changes would be 
represented by insertions on the internal black-hole propagator.
These insertions would couple together different black-hole
states, so that
\begin{eqnarray}
{\delta_{mn}\over s-M^2_m+i\Gamma_m} &\to& 
\sum_k {\delta_{mk}\over s-M^2_m+i\Gamma_m} \lambda_{kl}(s) 
{\delta_{ln}\over s-M^2_l+i\Gamma_l} + \ldots 
\nonumber \\
&=& \left({1\over s-M^2-\lambda(s)+i\Gamma}\right)_{mn} ,
\label{prop}
\end{eqnarray}
where $\lambda_{mn}(s)$ is a schematic representation of how 
a small change in the metric (or other external field) 
couples to the black-hole states.
Since these states are separated by an exponentially
small mass gaps, even very small $\lambda$'s have
a significant effect at large enough $s$.
Note that this effect is 
different from that of external fields that simply shift the
value of $s$ itself 
(such as tidal effects of the moon at the LEP accelerator \cite{lep});
in order to understand 
and compensate for eq.~(\ref{prop}), we would have to be able
to calculate the $\lambda_{mn}(s)$ matrices from first principles,
which in turn would require a detailed knowledge of the black
hole states.

\section{Implications for Experimental Measurements}
\label{4}

We begin by noting the obvious fact that nothing can
be learned about information loss by watching the
formation and evaporation of a {\it single\/} black
hole, even if it were possible to have perfect
measurements of the energies and spins of every
incoming and outgoing particle.  As always in
quantum mechanics, we must do repeated experiments
with ``identically prepared'' initial states
in order to measure cross sections, and from these
infer results about the $S$-matrix itself.
Thus our goal would be to measure scattering cross
sections precisely enough to discover the presence
and locations of the black-hole resonance poles,
eq.~(\ref{sn}).  This would require extraordinarily
precise measurements of both energy and event rate,
but there does not seem to be any obvious obstacle
to making these exacting measurements {\it in principle}.  

However, in the previous section we conjectured that
the precise resonance locations would be highly
sensitive to large-scale boundary conditions,
and therefore to the entire history of the past light-cone
at the spacetime location of the scattering event.
This history necessarily differs from event to event,
and so (we conjecture) do the locations of the resonance
poles.  Thus, while any {\it single\/} scattering event
could be described by a unitary $S$-matrix that maps
an initial density matrix $\rhoin$ to a final density
matrix $\rhout$,
\begin{equation}
\rhout = S \rhoin S^\dagger ,
\label{rhout}
\end{equation}
experimentally measured cross sections (which
necessarily require many scattering events) would
instead yield
\begin{equation}
\rhout = \sum_\gamma p_\gamma S_\gamma \rhoin S^\dagger_\gamma .
\label{rhout2}
\end{equation}
Here $\gamma$ stands schematically for the values of
the $C_n$'s and $\Gamma_n$'s (which change from event to
event), and $p_\gamma$ is a 
suitable probability distribution.  The theory of
random scattering matrices of this type is well
developed in mesoscopic physics, and is routinely
used to treat scatterers with chaotic 
properties (see, e.g., \cite{meso}).

In the present context, eq.~(\ref{rhout2}) is Hawking's
proposed ``dollar'' matrix that turns a pure
initial state into a mixed final state \cite{h1}.  
Thus, although the time evolution of the quantum state
of the universe is unitary, this fact cannot be
determined by making a series of repeated measurements
that are sensitive to the black-hole poles in the $S$-matrix.

If unitarity cannot be verified experimentally, then
it may be possible to devise a different, physically
equivalent, mathematical description in which density
matrices are the fundamental objects, the theory has
manifest field-theoretic locality, and non-unitary
evolution consistent with eq.~(\ref{rhout2}).  This
theory, if it exists,
would be experimentally indistinguishable
from the non-local, unitary version of physics that
results in eq.~(\ref{rhout}).

\section{Conclusions}
\label{5}
This paper has consisted of a series of increasingly bold
speculations, with the goal of producing a resolution, 
acceptable to all parties, of 
the debate of the last quarter-century concerning
the possible loss of quantum information in the evaporation
of black holes:  everyone is right.

A precise conjecture is that the locations of black-hole
resonances [that is, the poles in the $S$-matrix that
follow the general pattern of eq.~(\ref{sn})]
will depend sensitively on large-scale boundary conditions,
such as the moduli of a spatial compactification torus. 
We may hope that such calculations will become feasible
in the near future.

\begin{acknowledgments}
I am grateful to Jim Hartle, Gary Horowitz and Joe Polchinksi
for discussions and critical comments.
This work was supported in part by NSF Grant PHY00-98395.
\end{acknowledgments}

\end{document}